\documentclass[twocolumn]{aastex631}

\usepackage{CJK}

\graphicspath{{./}{figures/}}

\received{XXX}
\revised{YYY}
\accepted{ZZZ}

\usepackage{graphicx}	
\usepackage{amsmath}	
\usepackage{amssymb}	
\usepackage[inline]{enumitem}
\usepackage{gensymb}
\usepackage{array}
\newcolumntype{P}[1]{>{\centering\arraybackslash}p{#1}}
\newcolumntype{M}[1]{>{\centering\arraybackslash}m{#1}}
\usepackage{multirow}
\usepackage{tabularx}

\usepackage{color}

\usepackage{color}

\shorttitle{Two Novel Hot Jupiter Formation Pathways}
\shortauthors{Stephan et al. 2024}

\begin{document}
\begin{CJK*}{UTF8}{gbsn}

\title{Two Novel Hot Jupiter Formation Pathways:
\\How White Dwarf Kicks Shape the Hot Jupiter Population}

\correspondingauthor{Alexander P. Stephan}
\email{alexander.stephan@vanderbilt.edu}

\author[0000-0001-8220-0548]{Alexander P. Stephan}
\affiliation{Department of Physics and Astronomy, Vanderbilt University, Nashville, TN 37235, USA}

\author[0000-0002-7595-6360]{David V. Martin}
\affiliation{Department of Physics and Astronomy, Tufts University, Medford, MA 02155. USA}

\author[0000-0002-9802-9279]{Smadar Naoz}
\affiliation{Department of Physics and Astronomy, University of California, Los Angeles, Los Angeles, CA 90095, USA}
\affiliation{Mani L. Bhaumik Institute for Theoretical Physics, University of California, Los Angeles, Los Angeles, CA 90095, USA }

\author[0009-0005-5932-9702]{Nathan R. Hughes}
\affiliation{Department of Astronomy, The Ohio State University, Columbus, OH 43210, USA}

\author[0000-0003-1247-9349]{Cheyanne Shariat}
\affiliation{Department of Astronomy, California Institute of Technology, Pasadena, CA 91125, USA }
\affiliation{Department of Physics and Astronomy, University of California, Los Angeles, Los Angeles, CA 90095, USA}



\begin{abstract}

The origin of Hot Jupiters (HJs) is disputed between a variety of {\it in situ} and {\it ex situ} formation scenarios. One of the early proposed {\it ex situ} scenarios was the Eccentric Kozai-Lidov (EKL) mechanism combined with tidal circularization, which can produce HJs with the aid of a stellar or planetary companion. However, observations have revealed a lack of stellar companions to HJs, which challenges the importance of the binary star-driven EKL plus tides scenario. In this work, we explore so far unaccounted-for stellar evolution effects on HJ formation, in particular the effect of white dwarf (WD) formation. {\it Gaia} observations have revealed that WDs often undergo a kick during formation, which can alter a binary's orbital configuration or even unbind it. Based on this WD kick, in this letter we propose and explore two novel HJ formation pathways: 1) HJs that are presently orbiting single stars, but were initially formed in a binary that was later unbound by a WD kick; 2) Binaries that survive the WD kick can trigger enhanced EKL oscillations and lead to 2nd generation HJ formation. We demonstrate that the majority of seemingly single HJs could have formed in binary star systems. As such, HJ formation in binaries via the EKL mechanism could be one of the dominant HJ formation pathways, and our results highlight that unaccounted-for stellar evolution effects, like WD formation, can obscure the actual origin of observed exoplanet populations.

\end{abstract}


\section{Introduction}\label{sec:intro}

Observations over recent decades have uncovered a rich diversity of thousands of exoplanets, including many systems with configurations very unlike our own solar system and orbiting stars of all evolutionary phases \citep[e.g.,][]{WolszczanFrail1992,Charpinet+2011,Johnson+2011A,Gettel+2012,Howard+2012,Vanderburg+2020}. One of the most peculiar and unexpected types of discovered exoplanets was that of Hot Jupiters (HJs), gas giants with orbital periods of only a few days. HJs may only exist around $\approx 1\%$ of stars \citep[e.g.,][]{Howard+2012,Beleznay+2022}, and, as they do not occur in our solar system, their existence has challenged long-standing theories of planet formation. Uncovering the history of HJs has thus been an important open question.

A large variety of competing and, sometimes, overlapping theories of HJ formation has been explored over the years \citep[see, for example, a review by][]{DawsonJohnson2018}. While theories of in situ formation exist \citep[e.g.,][]{Batygin16,Boley+2016,BaileyBatygin18,Poon+21}, overall ex-situ theories appear to dominate the literature (i.e., the planet initially formed on a wider orbit and migrated inwards later in the system's evolution) due to the hostile conditions for large planet formation close to the star. These migration theories generally fall into two categories: smooth, disc-driven migration of a planet early in the system's evolution when the proto-planetary disc has not dissipated yet \citep[e.g.,][]{GoldreichTremaine1980,Lin+86,Lin+1996,Bodenheimer+2000,Mass+03,Heller2019,Mendigutia+24}, or tidally driven migration after the planetary orbit has been excited to high eccentricities \citep[e.g.,][]{Nag+08,Jackson+2008}. There is a variety of possible dynamical processes that could lead to these high orbital eccentricities, such as planet-planet scattering \citep[e.g.,][]{RasioFord1996,Chatterjee+2008}, secular chaos \citep[e.g.,][]{WuLithwick2011,Teyssandier+2019}, and the so-called Eccentric Kozai-Lidov (EKL) mechanism \citep{Kozai,Lidov,Naoz2016}. The EKL mechanism requires an additional distant companion to the host star and gas giant, which can be either another planet or a binary star companion. The EKL mechanism in stellar binaries and its potential for producing HJs has been of particular interest for many decades \citep[e.g.,][]{Mazeh+79,Innanen+1997,Hol+97,Wu+03,Fabrycky+07,Naoz+12bin,Stephan+2018,Weldon+24}.

\begin{figure*}[htbp]
    \centering
    \includegraphics[width=\linewidth]{pathways_cartoon.pdf}
    \caption{{\bf Cartoon diagrams of our two proposed novel formation pathways for Hot Jupiters.}\\ {\it Pathway 1 (top):} A cold Jupiter is turned into a HJ through EKL oscillations under the influence of a misaligned and more massive outer stellar companion. The more massive star evolves into a WD and receives a large velocity kick due to asymmetric mass loss. This kick ionizes the binary, leaving behind a HJ orbiting a single star.\\ {\it Pathway 2 (bottom):} A cold Jupiter is in a stellar binary where no EKL oscillations occur because the binary is not sufficiently misaligned with respect to the planet. The outer star, which is more massive, evolves into a WD. It receives a moderate velocity kick that misaligns the binary but does not ionize it, which in turn allows EKL oscillations to commence and leads to HJ formation. }
    \label{fig:pathways_cartoon}
\end{figure*}

The case for EKL-driven high-eccentricity tidal migration was strengthened by observations, such as the generally significant spin-orbit misalignments for HJs around hot stars vs.~the observed high degree of alignment around cool stars \citep[e.g.,][]{Tri+10,Winn+10b,Albrecht12,Albrecht:2022}. This precipitated many searches for stellar companions to HJs, (e.g.,\citealt{Knutson+2014,Ngo+2016,Bryan+2016}; review in \citealt{Martin2018}). While the observed binarity rate for HJ hosting systems is significantly higher than for non-HJ containing systems, some works in the literature have argued that the EKL mechanism cannot be a dominant cause of HJ formation, partially due to the high prevalence of single stars with HJs \citep[e.g.,][]{Petrovich2015a,Ngo+2016,MoeKratter21}. EKL-driven migration induced by a planetary companion has also been investigated \citep[e.g.,][]{Naoz11,Petrovich2015b} and recent observations of long-period giant planet companions to HJs have provided support for the idea \citep{ZinkHoward2023,DongHongWu+2023}.

Overall, the current literature appears to disfavor stellar binary EKL oscillations as the formation mechanism for more than a modest $\sim20~\%$ fraction of HJs, and instead promotes contributions from many different mechanisms to comparative degrees. However, here, 
we challenge the notion that binaries are only a minor site of HJ formation and show instead that stellar binary EKL oscillations do contribute a very large, potentially even dominant, fraction of HJs, which is a result of combining stellar evolution with the EKL mechanism. 

For a HJ that orbits the less massive companion star in a binary, the more massive, ``primary'' star will evolve off the main sequence (MS) before the planet host star, becoming a White Dwarf (WD) in most cases (if $<8M_\odot$). This configuration, a MS star with a HJ and an outer WD companion, has been found roughly a dozen times so far, as cataloged by \citet{Martin+21} with the recent discovery of TOI-1259. As a star evolves past the MS to become a WD, it can lose a significant amount of mass (on the order of a few percent to over $80~\%$), which can significantly alter the orbital properties of a binary system with planets \citep[e.g.,][]{Veras+2013,Veras2016,Stephan+2017,Stephan+2021} or a multi-planet system \citep[e.g.,][]{DebesSigurdsson2002, PetrovichMunoz2017}. See also reviews by \citet{Veras2016, Veras+2024}. Furthermore, it has been long suspected that this mass loss is slightly asymmetric near the end of the asymptotic giant branch (AGB) phase, which would induce a velocity recoil, commonly called a ``kick'' \citep{Spruit2002}. An observed dearth of WDs in open clusters provided early evidence for the idea \citep{Fellhauer2003,Heyl2008}. More recent evidence has come from the orbital parameters and occurrence rates of MS-WD and WD-WD binaries \citep{ElBadryRix18} and triples \citep{Shariat+2023} observed using {\it Gaia}. Furthermore, many close (few AU) MS-WD binaries, which presumably experienced an episode of mass transfer, are non-circular \citep[e.g.,][]{Lagos+2022, Shahaf+2024, Yamaguchi+2024b}, which may further support the presence of a kick-like mechanism associated with WD formation \citep[e.g.,][]{Shariat+2024}. The kick inferred from these observations is strong enough to significantly alter the orbital parameters of a stellar binary. This could in turn alter the dynamical evolution of the planet, potentially triggering new or stronger EKL-oscillations, leading to high-eccentricity tidal migration and HJ formation, or planetary engulfment. Furthermore, the kick is also strong enough to completely unbind about $\sim21~\%$ of all binaries, and the majority of binaries with orbits wider than about $1000$~AU \citep[][]{ElBadryRix18}.

In this paper, we demonstrate the significant effect that WD kicks have on the observed HJ population. Specifically, we propose two new HJ formation pathways that, to our knowledge, are presently absent in the literature:

\begin{enumerate}
    \item \textit{HJs presently orbiting single stars that used to be in a binary.} The HJ was formed in a stellar binary, potentially (though not necessarily) through stellar binary-induced EKL oscillations plus tidal migration and circularization. The outer (more massive) star evolved into a WD and experienced a formation kick from asymmetric mass loss and was ionized from the binary. The end result is an HJ that is observed orbiting a single star while it was originally formed in a binary. This is illustrated in Fig.~\ref{fig:pathways_cartoon} (top).
    \item \textit{Second-generation HJ formation in binaries due to WD kicks}. Initially, there is a cold Jupiter orbiting the less massive star of a MS-MS binary. Due to the initial orientation of the system (i.e., low mutual binary-planet inclination or small eccentricities), EKL oscillations are weak or non-existent. However, when the outer, more massive star evolves into a WD, it receives a kick that significantly changes the system configuration (in particular by raising the mutual inclination and/or eccentricities) but does not ionize the binary. EKL oscillations are thus enhanced enough to allow HJ formation via tidal migration. The end result is a HJ orbiting a MS star with an outer WD companion (similar to TOI-1259, \citealt{Martin+21}). This is illustrated in Fig.~\ref{fig:pathways_cartoon} (bottom).
\end{enumerate}

Overall, the effect of WD kicks is to obscure the evolutionary history of HJs. It is perilous to compare observed HJ populations in single and binary star systems, because a) some of the single stars either used to be binaries and the outer star was ionized, or b) some of the single stars are actually binaries with an outer WD companion that has been too faint to observe.

In fact, here we demonstrate that more HJs were formed in stellar binaries than previously thought. This could make binary-driven EKL oscillations and tidal migration the \textit{dominant} HJ formation mechanism. These two novel HJ formation pathways that we describe in this work could thus challenge and alter the accepted formation history statistics of the overall HJ population. The details of our model and simulation parameters are outlined in Section \ref{sec:methods}, with the results of our calculations presented in Section \ref{sec:res} and their implications discussed in Section \ref{sec:disc}. We summarize our work and present our overall conclusions in Section \ref{sec:summary}.

\section{Numerical Methods and Simulation Parameters}\label{sec:methods}

The basic system architecture under consideration here consists of a wide binary star with one of the two stars being orbited by a giant planet. For such systems to be long-term stable (on the time scale necessary for stellar evolution), they generally need to be hierarchical, meaning that, over short time scales, the dynamics can be described by an inner and outer Keplerian orbit with semi-major axes (SMAs) $a_1$ and $a_2$, respectively, inner and outer eccentricity $e_1$ and $e_2$, and mutual inclination $inc$. To ensure orbital stability and a hierarchical system, generally $a_1/a_2 < 0.1$ and $\epsilon < 0.1$, with $\epsilon$ being defined as \begin{equation}
    \epsilon = \frac{a_1}{a_2}\frac{e_2}{1 - e_2^2} \ ,
\end{equation} \citep[see, for example,][]{Naoz2016}. Over timescales much longer than the orbital timescales the inner and outer orbits will interact by exchanging angular momenta, the effects of which have long been studied as the so-called Kozai-Lidov (KL) effect \citep{Kozai,Lidov}, and which, in more recent years, has been expanded to the Eccentric Kozai-Lidov (EKL) mechanism \citep[e.g.,][]{Naoz+11sec}. In this case, the system evolution can be treated secularly, following the orbit-averaged equations from solving the hierarchical three-body Hamiltonian up to the octupole order of approximation. Our secular calculations also include the long-term dynamical effects of equilibrium tides \citep[following][assuming viscous timescales of $1.5$~years for both stars and gas giant planets]{Hut,1998EKH}, the aforementioned stellar evolution effects of mass loss and radius expansion \citep[following the stellar evolution code {\tt SSE} by][]{Hurley+00}, as well as post-Newtonian corrections from general relativity \citep{Naoz+12GR}. The full set of equations is also given by \citet{Naoz2016}. The combination of these physical processes has been well tested by us in a variety of previous dynamical studies \citep[e.g.,][]{Stephan+2016,Stephan+2017,Stephan+2018,Stephan+2019,Stephan+2021,Angelo+2022,Shariat+2023,Shariat+2024}.

\begin{figure*}[htbp]
    \centering
    \includegraphics[width=\linewidth]{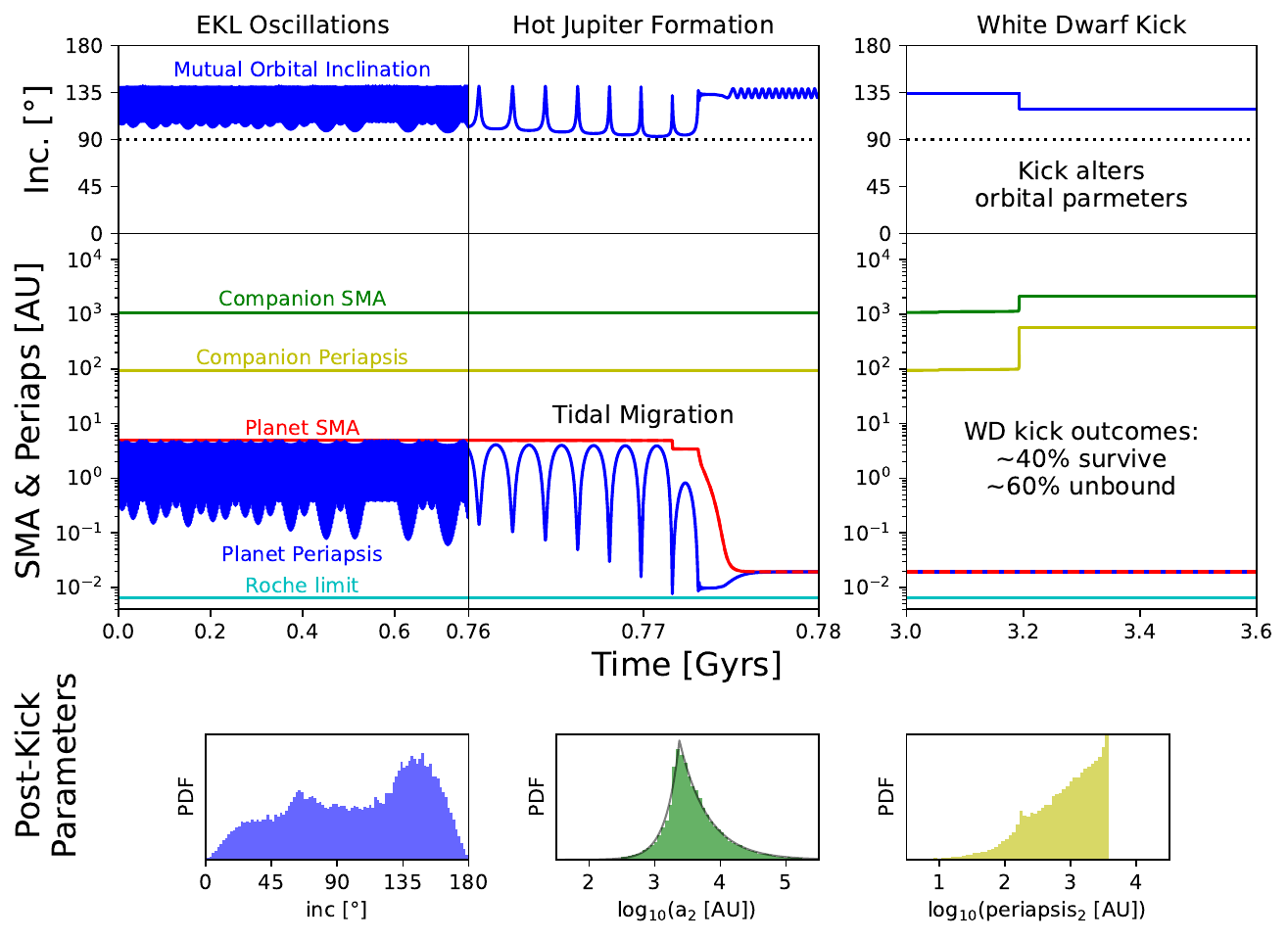}
    \caption{{\bf Evolution of a Binary Star System with a WD Formation Kick post Hot Jupiter Formation.} Shown is the evolution of a Jupiter-like planet orbiting in a binary star system and the outcomes of the WD formation kick. The upper row of panels shows the mutual inclination between the planetary and stellar binary orbits. In the middle row of panels, the red and blue lines show the planet's SMA and periapsis distance, respectively; the cyan line shows the planet's Roche limit and the green and yellow lines show the stellar companion's SMA and periapsis distance, respectively. Note that the planet undergoes eccentricity oscillations due to the Eccentric Kozai-Lidov effect, leading to tidal migration and circularization as a Hot Jupiter at around $0.77$~Gyrs into the system's evolution. We determine that tidal decay after circularization is minimal for the planet, such that it will approximately remain in its orbit from that point onwards. At around $3.2$~Gyrs, the more massive stellar companion evolves off the MS, becoming first a red giant and finally a WD. The mass loss during that process results in an expansion of the companion's SMA and periapsis distance. Finally, we include the effects of the WD formation kick following \citet{ElBadryRix18}, which results in a new mutual orbital inclination and new companion SMA and periapsis in $\sim40\%$ of cases, which, for this particular system, follow the blue, green, and yellow distributions in the three plots on the bottom of the figure, respectively. The stellar binary is separated in $\sim60\%$ of cases for this system. The right column of panels shows one random example of surviving system parameters. The planet host star has an initial mass of m$_1=0.516$~M$_{\odot}$, the planet has the mass and radius of Jupiter and an initial orbit SMA of a$_1=4.935$~AU and eccentricity of e$_1=0.01$, the stellar companion has an initial mass of m$_3=1.486$~M$_{\odot}$, initial orbit SMA of a$_2=1072.766$~AU and eccentricity of e$_2=0.913$. The planetary and stellar orbits have an initial mutual inclination of inc~$=107.091\degree$ and arguments of periapsis of $\omega_1 = 111.704\degree$ and $\omega_2 = 252.729\degree$, respectively. The stellar radii and mass loss are calculated by {\tt SSE} throughout the evolution.}
    \label{fig:ExampleEvo1}
\end{figure*}

\begin{figure*}[htbp]
    \centering
    \includegraphics[width=\linewidth]{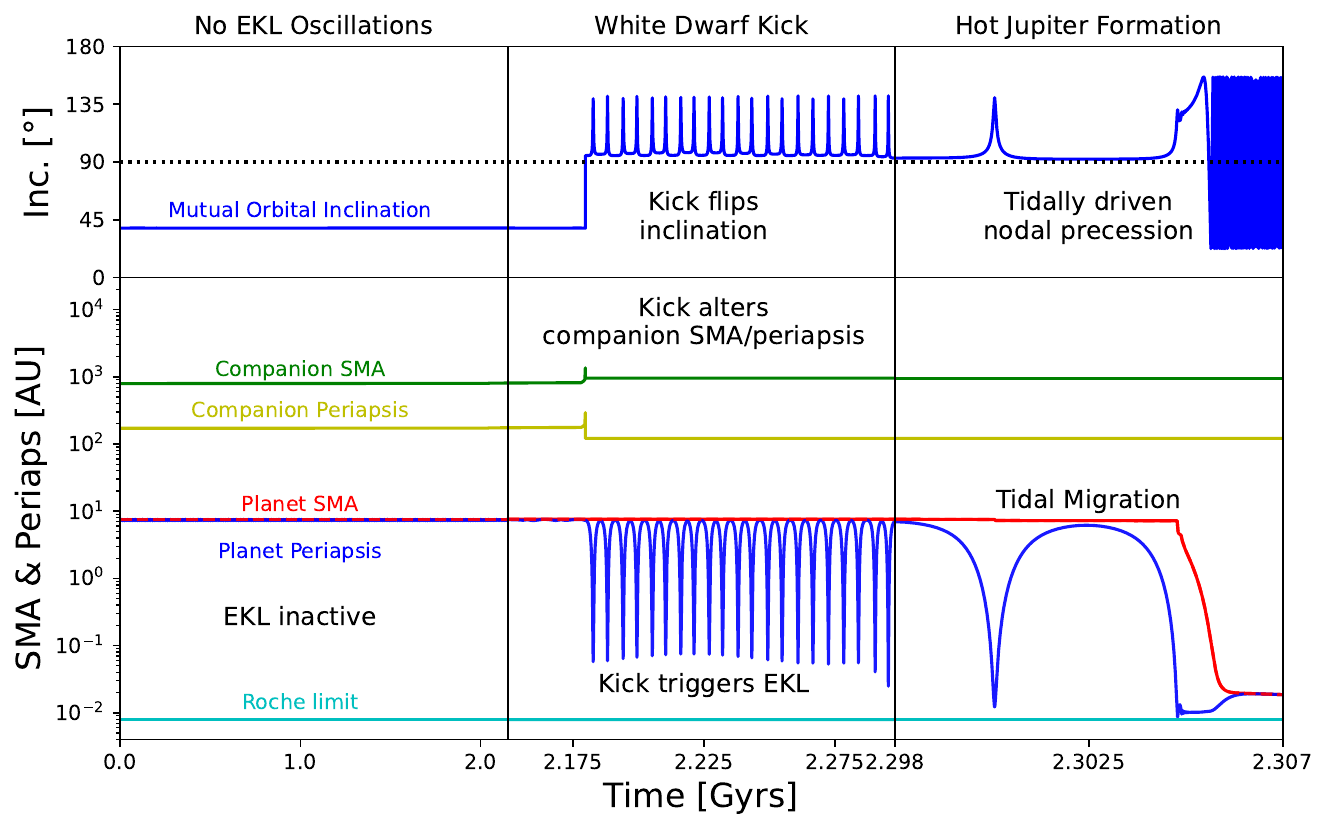}
    \caption{{\bf Evolution of a Binary Star System that forms a Hot Jupiter due to a WD Formation Kick.} The line colors used in this plot correspond to the same parameters as shown in Fig.~\ref{fig:ExampleEvo1}. Unlike the system shown in Fig.~\ref{fig:ExampleEvo1}, the system shown here does not undergo KL oscillations at the beginning (left column of panels), as the mutual inclination of the planetary and stellar binary orbits is too small. However, the WD formation kick (middle column of panels) is able to alter the companion inclination, SMA, and eccentricity sufficiently to trigger KL evolution, which eventually leads to Hot Jupiter formation (right column of panels). The planet host star has an initial mass of m$_1=0.930$~M$_{\odot}$, the planet has the mass and radius of Jupiter and an initial orbit SMA of a$_1=7.583$~AU and eccentricity of e$_1=0.01$, the stellar companion has an initial mass of m$_3=1.683$~M$_{\odot}$, initial orbit SMA of a$_2=791.617$~AU and eccentricity of e$_2=0.784$. The planetary and stellar orbits have an initial mutual inclination of inc~$=38.523\degree$ and arguments of periapsis of $\omega_1 = 64.291\degree$ and $\omega_2 = 90.128\degree$, respectively.}
    \label{fig:ExampleEvo2}
\end{figure*}

In addition to the processes mentioned above, we also include kicks and the effects of sudden mass loss to the WD when it forms at the end of the asymptotic giant branch (AGB) phase. To incorporate such non-adiabatic events into a secular three-body calculation, we follow the formalism described by \citet{LuNaoz2019}\footnote{See also, \citet{Hamers18,Hoang+22,Jurado+24}.}, which has been tested for WD formation kicks in recent works \citep{Shariat+2023,Shariat+2024}. As our calculations are secular, the particular positions of the stars and the planets in their orbits are chosen randomly at the time of the kick, and new orbital parameters are determined before returning to the secular calculations. The kick direction is chosen randomly, and the kick velocity $v_{\rm kick}$ is picked from the Maxwellian distribution empirically determined by \citet{ElBadryRix18}, with the probability function
\begin{equation}
    P(v_{\rm kick}) = \sqrt{\frac{2}{\pi}}\frac{v_{\rm kick}^2}{\sigma_{\rm kick}^3}\exp{\left[-\frac{v_{\rm kick}^2}{2\sigma_{\rm kick}^2}\right]} \ ,
\end{equation} 
with $\sigma_{\rm kick}=0.5$~km~s$^{-1}$ and with the peak of the distribution at $v_{\rm kick}=\sqrt{2}\sigma_{\rm kick}\approx0.7$~km~s$^{-1}$. Examples of systems following the dynamical evolution and kick prescription outlined here are shown in Fig.~\ref{fig:ExampleEvo1} and Fig.~\ref{fig:ExampleEvo2}. We note here that we are agnostic about the exact nature of the kick mechanism, as it has not yet been clearly determined in the literature, As such, we assume the kick happens instantaneously at the very end of the AGB phase, after most gradual mass loss from stellar winds has already occurred. We also tested the consistency of including the kicks in our secular calculations by comparing them with numerical calculations using the N-body code {\tt REBOUND} and its library {\tt REBOUNDx} \citep{ReinLiu2012,Tamayo+2020}.

Using the equations and effects outlined above, we calculate the secular evolution of a total of $12,823$ systems. The mass of the more massive companion star (technically the ``primary'' star of the system) is drawn from a Salpeter distribution with a minimum mass of $1.1$~M$_{\odot}$ and a maximum mass of $8.0$~M$_{\odot}$, to ensure that the star will evolve off the MS within the age of the universe and does not undergo a supernova. The choice of this mass range is further justified as the likelihood of a star being in a binary increases rapidly with mass \citep{Raghavan+10}, with most solar mass or heavier stars residing in binaries or even higher multiples. The mass of the less massive planet host star is drawn from the mass ratio distribution determined by \citet{Duquennoy+91}, with a minimum mass of $0.1$~M$_{\odot}$ to ensure it will indeed be a star. The planet has the mass, radius, and spin of Jupiter. The planet's SMA is drawn from a uniform distribution between $1$ and $10$~AU, and its initial eccentricity is set to $0.01$. The companion star's SMA is drawn from the log-normal binary period distribution also determined by \citet{Duquennoy+91}, and its eccentricity is drawn from a uniform distribution between $0$ and $1$, tested for orbital stability following the criteria outlined before. The initial arguments of periapsis are drawn uniformly for both orbits between $0\degree$ and $360\degree$. The initial mutual inclination between the orbits is drawn isotropically. This leads to about $77\%$ of our systems starting with inclinations between $\sim40\degree$ and $\sim140\degree$, which is the parameter space where the quadrupole level of approximation yields large eccentricity excitations \citep[e.g.,][]{Li+14,Hansen+20}. 
The systems are evolved for $13$~Gyrs, or until one of our stopping conditions is fulfilled. These stopping conditions include 1) Planet engulfment or Roche lobe crossing; 2) Planet orbital circularization after tidal migration (HJ formation); 3) Unbinding of the binary from WD formation kick. For systems that form HJs before the companion evolves into a WD, we check the binary stability against kicks by testing $1000$ different randomly drawn kicks from the \citet{ElBadryRix18} distribution at the end of their RGB lifetimes.

\section{Results}\label{sec:res}

Our $12,823$ simulations produce overall a total of $245$ HJs, $162$ of which become HJs before the companion becomes a WD and undergoes a kick (see Fig.~\ref{fig:ExampleEvo1} for an example system evolution), $83$ of which become HJs after and due to the WD kick (see Fig.~\ref{fig:ExampleEvo2}). $3810$ planets become engulfed by their host stars ($2267$ before kick, $1543$ after and due to the kick), $4741$ systems do not form HJs, and their binaries are disrupted via the WD formation kick, and $3665$ never have strong star-planet interactions and survive the kick. $362$ of our runs never reach a definitive end-state within a reasonable computation time due to various factors, such as violating the stability and hierarchicalness conditions, and we discard them for our analysis rather than to speculate on the eventual system evolution result. See Table \ref{tab:table1} for a full list of outcomes.

\begin{table*}[t]
 \begin{center}
  \caption{Outcomes of Binary with cold Jupiter Simulations.}
  \label{tab:table1}
  \begin{tabular}{|c||c|c|c|c|}
        \hline
        \multicolumn{5}{|c|}{Initial Binaries with cold Jupiters: 12823} \\
        \hline
        Binary Phase & HJ formation & Planet Engulfment & Kick Separation & Incomplete \\
        \hline
        \hline
        All & 245 & 3810 & 4741 & 362 \\
        \hline
        MS-MS & 162 & 2267 & - & - \\
        \hline
        MS-WD & 83 & 1543 & 4741 & -\\
        \hline
        \multicolumn{5}{|c|}{Remaining Binaries with cold Jupiters: 3665} \\
        \hline
  \end{tabular}     
 \end{center}
\end{table*}

\begin{figure}[htbp]
    \centering
    \includegraphics[width=\linewidth]{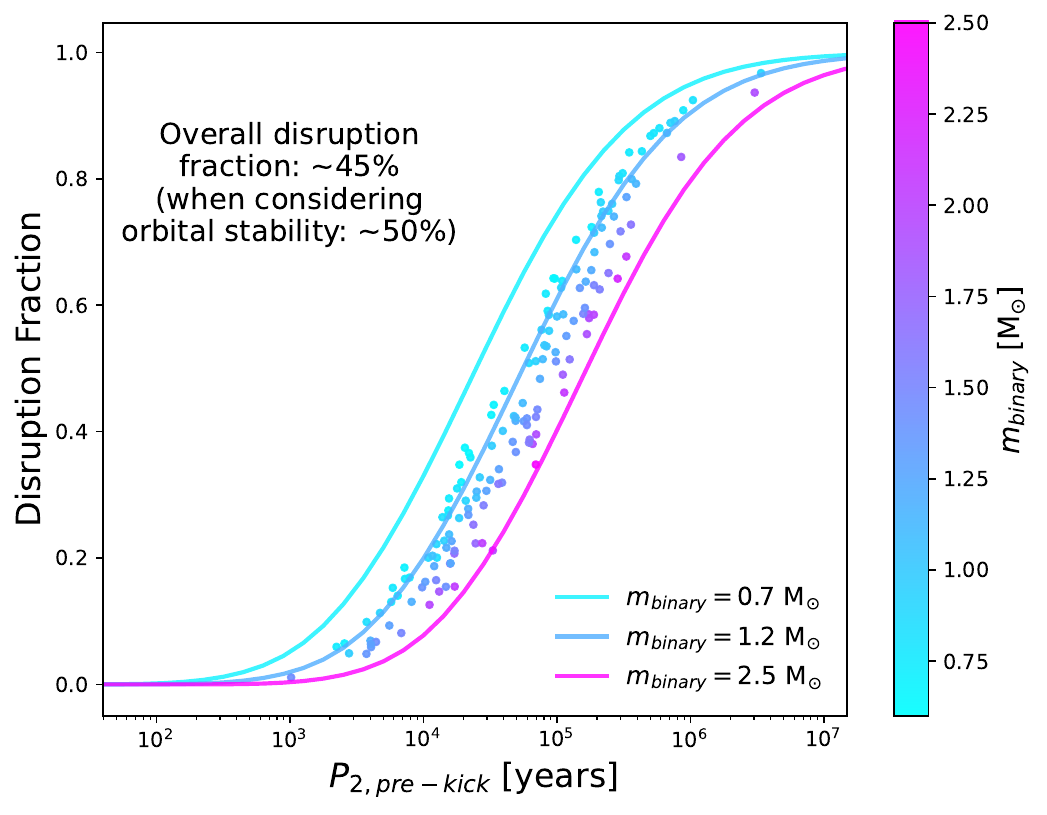}
    \caption{{\bf Likelihood of binary disruption vs. pre-kick stellar binary period for HJ forming systems in our simulations.} Shown is the likelihood that a particular HJ forming system in our simulations is disrupted by the WD formation kick versus the orbital period of the stellar binary, $P_2$, just before WD formation, color coded by the mass of the stellar binary. The cyan, blue, and magenta lines show the disruption likelihood versus orbital period for stellar binaries with eccentricity $0.6$ and with total masses of $0.7$, $1.2$, and $2.5$~M$_{\odot}$, respectively. Note that these are the masses towards the end of the AGB, when some significant mass loss has already occurred. Overall, our sample of HJ systems will undergo binary disruption from the WD formation kick in about $45\%$ of cases, about $50\%$ when considering the orbital stability of the post-kick system.}
    \label{fig:Disrupt_vs_P2}
\end{figure}

The $162$ HJs that form prior to their stellar binary companions' WD kick are roughly $1.26~\%$ of our binary sample, which is a formation efficiency that is overall consistent with stellar binary KL induced HJ formation efficiencies of various previous studies \citep[e.g.,][]{Naoz+12bin,Petrovich2015a,Anderson+2016,Stephan+2018}. We use this sample of HJs and test the likelihood for their binary systems to be disrupted by the WD formation kick as described in Section \ref{sec:methods}. The likelihood of disruption versus the binaries' orbital period before the kick is shown in Fig.~\ref{fig:Disrupt_vs_P2}. Overall, we see that about $45~\%$ of the pre-WD binaries that have formed HJs in our simulations will be disrupted by the kicks, with an additional $5~\%$ resulting in unstable configurations (i.e., the binary periapsis is inside or too close to the planet's orbit, or the new binary SMA is so wide that galactic tides and stellar flybys quickly lead to disruption), implying a $50~\%$ binary disruption rate from WD kicks. The two dominating factors that determine disruption likelihood are a binary's pre-kick mass and orbital separation or period, with orbital eccentricity being a minor additional factor. Since we are only considering HJs orbiting the less massive binary companion, the overall disruption rate has to be reduced by a factor of $2$, implying that $25~\%$ of HJs in binaries will lose the companion due to the WD kick.

The $83$ HJs that form due to the WD kick's alteration of the binary orbital parameters constitute roughly $0.65~\%$ of our total binary sample. However, when considering that only $5,653$ of our binaries actually reach the post-kick phase without pre-kick HJ formation, planet engulfment, or binary separation, the post-kick HJ formation efficiency is about $1.47~\%$, somewhat higher but comparable to the pre-kick HJ formation efficiency mentioned above. As such, our simulations imply that HJs in binaries with WD companions have about an equal chance to have become HJs before or after the companion became a WD.

\section{Discussion}\label{sec:disc}

By including WD formation kicks in the dynamics of Jupiter-like planets in stellar binaries, our simulations have revealed two distinct evolutionary pathways of HJ formation that may have a significant impact on our overall understanding of the observed HJ population, which we discuss here: 1) ``lonely'' HJs orbiting single stars that are the remnants of kick-disrupted binaries (Fig.~\ref{fig:pathways_cartoon}, top), and 2) ``second generation'' HJs that have formed due to the binary orbital changes induced by the kick altering the dynamical behavior of the system (Fig.~\ref{fig:pathways_cartoon}, bottom). Our results indicate that these types of HJs could constitute on the order of about half of the HJ population in the binary-triggered high-eccentricity migration framework, which may require significant adjustments to statistical models of the HJ population.

\subsection{Single Star HJs from Binary Separations}\label{subsec:singleHJs}

Our results in Sec.~\ref{sec:res} show that about a quarter of all HJs that form via KL-induced high-eccentricity migration in MS stellar binaries will lose their stellar companion via the WD formation kick. After kick disruption, these HJs will practically look indistinguishable from other ``lonely'' HJs to an observer. Any population-level observational study investigating the various possible formation origins of HJs would most likely consider a non-binary or non-dynamical formation history for these objects when observed, which would bias the conclusions drawn from the population statistics against dynamical origins of HJ formation. We can briefly estimate the potential impact of this bias by considering the observed stellar companion rates to HJs. 

Previous studies have generally found that about $50~\%$ of observed HJs orbiting MS stars have main-sequence stellar companions out to separations of about $2,000$~AU \citep[e.g.,][]{Ngo+2016}. Assuming that all HJs in these binaries are produced via stellar binary KL, and given our disruption chance of $25~\%$, the initial, pre-kick fraction of HJs in binaries would be closer to $66.7~\%$, or $2/3$ of the HJ population, such that the observed post-kick fraction is only $50~\%$. However, we can also consider that most HJ observational surveys targeting binaries will only search for HJs orbiting the brighter, primary star \citep[e.g.,][]{Ngo+2016}, which will further bias the statistics. Specifically, if the $50~\%$ of HJs with binaries are assumed to all be HJs around the primary stars, but the $50~\%$ of HJs without any observed companions can also come from disrupted binaries where the HJ formed around the less massive secondary star, then half of the observed HJs with no companions should originate from disrupted binaries. \textit{Extrapolating to the overall HJ population, this would imply that up to $80~\%$ of HJs originate in binary systems.}\footnote{The $80~\%$ estimate is also consistent with some estimates based on the observed HJ obliquity distribution \citep[e.g.,][]{Naoz+12bin}. } As such, a sensible estimate for the true rate of HJs forming in stellar binaries should lie between $66.7~\%$ and $80~\%$, implying strongly that stellar EKL-induced high-eccentricity migration is a dominant HJ formation mechanism. Furthermore, the observational statistics of HJs in binaries that are wider than $2,000$~AU are poorly understood, and many observed HJs in singles may have hidden binary companions (such as cool WDs on wide orbits, as discussed in Sec.~\ref{subsec:2ndGen}), which may further suppress the observed fraction of HJs with stellar companions. \textit{It is, therefore, possible that virtually all observed ``lonely'' HJs formed originally in stellar binaries, after which they simply lost their stellar companions, or they have stellar remnant companions that have eluded detection.} These numbers, of course, do not consider HJs in multi-planet systems \citep[which may be about $50~\%$ of HJs according to ][]{Knutson+2014}, which should form via planet-induced KL or other dynamical pathways.

\subsection{Second Generation HJs}\label{subsec:2ndGen}

Our results suggest that the formation of ``second generation'' HJs due to the WD kick is a robust process that could produce about $1/3$ of all EKL-induced HJs in stellar binaries. In fact, we estimate the number of second-generation HJs to be about the same as the number of pre-kick HJs that lose the stellar companion due to the kick, and as the number of pre-kick HJs that retain their stellar companion after the kick in our simulations (not considering HJs around the primary stars). This has several interesting implications for HJ observational statistics.

Firstly, for any HJ that is found to have a WD companion, the chances that the HJ formed before the WD kick or after and due to the WD kick are about the same in the high-eccentricity migration framework. The study of HJs in such systems may thus be of particular interest, as half of them will have become HJs only relatively recently in their home system's evolutionary history, with the cooling age of the WD providing a limit for its age as a HJ, and their atmospheric chemistry and internal temperature may differ from that of pre-kick HJs that have been HJs for many billions of years already. 

Secondly, binary systems that are comprised of a MS star and a WD companion are effectively over twice as likely to have an HJ around the MS star than around either star in a regular MS-MS star binary. Even when considering that either MS star in a MS-MS binary could have an HJ, the chances for a MS-WD system to have an HJ is still larger than the chance for a MS-MS system \citep[especially since WDs are also able to have HJs, as observed and explored by][]{Vanderburg+2020,OConnor+2020,Munoz+2020,Stephan+2021}. As more and more MS-WD binaries are found thanks to {\it Gaia} and other surveys, these types of systems should thus be a rich source of future HJ discoveries. Some work has already been done in this regard, showing an ever-growing sample of HJs in MS-WD systems \citep[e.g.][]{Martin+21}.

Lastly, the prevalence of HJs in MS-WD systems, both from pre-kick and post-kick formation channels, contributes to the observational bias against binaries as the evolutionary cause of HJ formation (see discussion in Sec.~\ref{subsec:singleHJs}), as WD companions to an MS star are generally more difficult to detect, especially once they have cooled down and start to dim \citep[see also introduction to][]{ElBadryRix18}. Thus, HJs that are believed to be in single MS star systems may indeed often have distant and cool WD companions.

\subsection{Caveats and Future Work}\label{subsec:caveats}

While our results clearly show that binary ionization is the {\it eventual} outcome of many stellar binaries containing HJs, it is important to consider for how long a given binary with HJ will be observable as a MS-MS binary vs.~as a MS-WD binary or kick-separated single star, in order to gauge the likelihood that observed lonely HJs were previously in binary systems. This depends on several factors, such as the masses of the binary star, the time of system formation, and the exact tidal efficiencies and, thus, tidal circularization times. Within our simulations, we find that when giving the binaries lifetimes of about the age of the universe of $13$~Gyrs, an average HJ from a separated binary will have spent about $12$ times longer orbiting the single star than orbiting within the original binary, starting from the time of orbit circularization. It is therefore much more likely to ``catch'' such systems post kick-disruption, however using a realistic star formation history may alter this conclusion. As such, we are aware that this could be an important caveat for our assessment that most lonely HJs started their existence in binary systems. This issue will be addressed in future work, where we plan to conduct a population synthesis study using a realistic star formation and binary formation history to deduce the overall evolution of the HJ population. 

A similar caveat presents itself concerning the survival lifetimes of HJs post orbital circularization, and the likelihood of tidal damping overcoming the EKL-oscillations that may drive the planet to stellar engulfment. The long-term orbital decay of HJs is a complex problem that relies on the exact tidal mechanisms at work in these systems, where our simplistic equilibrium tide model may be insufficient and more complex mechanisms like dynamical, obliquity, resonant, or chaotic tides may need to be considered \citep[e.g.,][]{Fabrycky+07,Anderson+2016,Teyssandier+2019,ZanazziWu2021}. For the purposes of this current work, we have generally assumed that HJs, once circularized around a MS star, will generally remain as HJs until engulfment as their host star expands when it reaches the RG and AGB phases. We have assumed viscous timescales for both stars and gas giant planets to be on the order of $1.5$~years \citep[see appendix of][for the details of the equilibrium tide model]{Naoz+2016}; a longer or shorter viscous timescale may alter our simulations' HJ formation efficiency (see Sec.~\ref{sec:res}).

We also acknowledge that WD kicks are not the only mechanism that can lead to binary disruption. Galactic tides and stellar encounters are known to shape the binary orbital parameter distribution and can lead to binary disruptions \citep[e.g.,][]{Binney+Tremaine1987,MichaelyShara2021,MichaelyNaoz2023,Yael+2022,MorPerets2023,HamiltonModak2024}. However, in the solar neighborhood, these mechanisms only appear to be of concern for binaries with SMAs of several thousand AU or wider, with a typical expected disruption timescale from fly-bys of about $3.8$~Gyrs for a $10,000$~AU wide binary, rising to $13$~Gyrs for a $3000$~AU wide binary \citep[see Eq.~16 of][]{HamiltonModak2024}. Virtually all our HJ-producing systems have binary SMAs below $3000$~AU, we therefore expect fly-bys and galactic tides to have minimal impact on our simulations' HJ formation efficiency. However, the often increased SMAs of the binaries post-WD formation kick will generally make MS-WD binaries more susceptible to disruption by these secondary effects, which we estimate to contribute additional disruptions on the order of a few percent (see Sec.~\ref{sec:res} and Fig.~\ref{fig:Disrupt_vs_P2}).

\section{Summary and Conclusions}\label{sec:summary}

In this letter 
we have presented two new evolutionary pathways for HJ formation that are the consequence of WD formation kicks: 1.~HJs around single stars that were formed in stellar binaries and have lost the binary companions via WD formation kicks; 2.~HJs in MS-WD binaries that have formed due to the WD kick changing the orbital parameters of the binary, triggering or enhancing the EKL mechanism and high eccentricity migration. The existence of these new pathways has several key consequences for our understanding of the observed HJ population as a whole:

\begin{enumerate}
    \item Given that about $50~\%$ of HJs are observed to have distant stellar binary companions and that generally only the primary star in a binary is investigated for finding HJs, we estimate that {at least $80~\%$ of HJs could have originated in binary systems}, ignoring HJs in multi-planet systems. As such, given also that finding distant WD companions to stars with HJs is challenging and has only recently been enabled by surveys conducted by {\it Gaia}, it is possible that all HJs orbiting single stars could have been formed in binary systems and have simply lost their companions via WD formation kicks.

    \item MS-WD binary systems could be more likely than MS-MS systems to contain HJs orbiting an MS star. As such, MS-WD binaries should be a primary focus for the search for new HJs and for finding interesting targets for planetary atmosphere observations, as these HJs will also have relatively well-defined ages in about $50~\%$ of cases. {\it TESS} and {\it Gaia} have already aided in finding an ever-growing list of such interesting systems, also including HJs around WDs.

    \item The overall contributions of various formation mechanisms to the overall HJ population need to be re-evaluated. Our results suggest that the contribution to the HJ population of the EKL mechanism in binary systems has been significantly underestimated, and that most, if not all, HJs could be produced via dynamical mechanisms such as the EKL mechanism, both stellar and planetary, and other planet-planet interactions such as scattering or resonances. As such, the contributions to the HJ population of ``in situ'' formation mechanisms may be much less than previously thought. In a future work we will conduct a population synthesis study to more accurately investigate the possible contribution of the various formation mechanisms when including the two new formation pathways explored in this work.
\end{enumerate}

The results of this work highlight the importance of various stellar evolution effects on the overall HJ population. Given that the vast majority of post-main sequence stars at the current age of the universe are WDs that have most likely experienced the WD formation kicks uncovered by \citet{ElBadryRix18} and that most current WDs have been in binary systems during the MS, this work is taking a crucial step towards completing the HJ formation pathway catalog.

\section{Acknowledgments}
A.P.S. acknowledges support from NASA grant 80NSSC24M0022. 
S.N. acknowledges partial support from NASA XRP grant 80NSSC23K0262 and thanks Howard and Astrid Preston for their generous support. 
N.R.H. acknowledges generous support by the Ohio State University's Department of Astronomy and Center for Cosmology and AstroParticle Physics via the Summer Undergraduate Research Program in Astrophysics (SURP).

\software{Matplotlib \citep{Hunter2007}, NumPy \citep{Harris+2020}, SciPy \citep{Virtanen+2020}, Rebound \citep{ReinLiu2012}, Reboundx \citep{Tamayo+2020}}





\bibliographystyle{aasjournal}
\bibliography{Kozai2}{} 

\begin{thebibliography}{}
\expandafter\ifx\csname natexlab\endcsname\relax\def\natexlab#1{#1}\fi
\providecommand{\url}[1]{\href{#1}{#1}}
\providecommand{\dodoi}[1]{doi:~\href{http://doi.org/#1}{\nolinkurl{#1}}}
\providecommand{\doeprint}[1]{\href{http://ascl.net/#1}{\nolinkurl{http://ascl.net/#1}}}
\providecommand{\doarXiv}[1]{\href{https://arxiv.org/abs/#1}{\nolinkurl{https://arxiv.org/abs/#1}}}

\bibitem[{{Albrecht} {et~al.}(2012){Albrecht}, {Winn}, {Johnson}, {Howard},
  {Marcy}, {Butler}, {Arriagada}, {Crane}, {Shectman}, {Thompson}, {Hirano},
  {Bakos}, \& {Hartman}}]{Albrecht12}
{Albrecht}, S., {Winn}, J.~N., {Johnson}, J.~A., {et~al.} 2012, \apj, 757, 18,
  \dodoi{10.1088/0004-637X/757/1/18}

\bibitem[{{Albrecht} {et~al.}(2022){Albrecht}, {Dawson}, \&
  {Winn}}]{Albrecht:2022}
{Albrecht}, S.~H., {Dawson}, R.~I., \& {Winn}, J.~N. 2022, \pasp, 134, 082001,
  \dodoi{10.1088/1538-3873/ac6c09}

\bibitem[{{Anderson} {et~al.}(2016){Anderson}, {Storch}, \&
  {Lai}}]{Anderson+2016}
{Anderson}, K.~R., {Storch}, N.~I., \& {Lai}, D. 2016, \mnras, 456, 3671,
  \dodoi{10.1093/mnras/stv2906}

\bibitem[{{Angelo} {et~al.}(2022){Angelo}, {Naoz}, {Petigura}, {MacDougall},
  {Stephan}, {Isaacson}, \& {Howard}}]{Angelo+2022}
{Angelo}, I., {Naoz}, S., {Petigura}, E., {et~al.} 2022, \aj, 163, 227,
  \dodoi{10.3847/1538-3881/ac6094}

\bibitem[{{Bailey} \& {Batygin}(2018)}]{BaileyBatygin18}
{Bailey}, E., \& {Batygin}, K. 2018, \apjl, 866, L2,
  \dodoi{10.3847/2041-8213/aade90}

\bibitem[{{Batygin} {et~al.}(2016){Batygin}, {Bodenheimer}, \&
  {Laughlin}}]{Batygin16}
{Batygin}, K., {Bodenheimer}, P.~H., \& {Laughlin}, G.~P. 2016, \apj, 829, 114,
  \dodoi{10.3847/0004-637X/829/2/114}

\bibitem[{{Beleznay} \& {Kunimoto}(2022)}]{Beleznay+2022}
{Beleznay}, M., \& {Kunimoto}, M. 2022, \mnras, 516, 75,
  \dodoi{10.1093/mnras/stac2179}

\bibitem[{{Binney} \& {Tremaine}(1987)}]{Binney+Tremaine1987}
{Binney}, J., \& {Tremaine}, S. 1987, {Galactic dynamics}

\bibitem[{{Bodenheimer} {et~al.}(2000){Bodenheimer}, {Hubickyj}, \&
  {Lissauer}}]{Bodenheimer+2000}
{Bodenheimer}, P., {Hubickyj}, O., \& {Lissauer}, J.~J. 2000, \icarus, 143, 2,
  \dodoi{10.1006/icar.1999.6246}

\bibitem[{{Boley} {et~al.}(2016){Boley}, {Granados Contreras}, \&
  {Gladman}}]{Boley+2016}
{Boley}, A.~C., {Granados Contreras}, A.~P., \& {Gladman}, B. 2016, \apjl, 817,
  L17, \dodoi{10.3847/2041-8205/817/2/L17}

\bibitem[{{Bryan} {et~al.}(2016){Bryan}, {Knutson}, {Howard}, {Ngo}, {Batygin},
  {Crepp}, {Fulton}, {Hinkley}, {Isaacson}, {Johnson}, {Marcy}, \&
  {Wright}}]{Bryan+2016}
{Bryan}, M.~L., {Knutson}, H.~A., {Howard}, A.~W., {et~al.} 2016, \apj, 821,
  89, \dodoi{10.3847/0004-637X/821/2/89}

\bibitem[{{Charpinet} {et~al.}(2011){Charpinet}, {Fontaine}, {Brassard},
  {Green}, {Van Grootel}, {Randall}, {Silvotti}, {Baran}, {{\O}stensen},
  {Kawaler}, \& {Telting}}]{Charpinet+2011}
{Charpinet}, S., {Fontaine}, G., {Brassard}, P., {et~al.} 2011, \nat, 480, 496,
  \dodoi{10.1038/nature10631}

\bibitem[{{Chatterjee} {et~al.}(2008){Chatterjee}, {Ford}, {Matsumura}, \&
  {Rasio}}]{Chatterjee+2008}
{Chatterjee}, S., {Ford}, E.~B., {Matsumura}, S., \& {Rasio}, F.~A. 2008, \apj,
  686, 580, \dodoi{10.1086/590227}

\bibitem[{{Dawson} \& {Johnson}(2018)}]{DawsonJohnson2018}
{Dawson}, R.~I., \& {Johnson}, J.~A. 2018, \araa, 56, 175,
  \dodoi{10.1146/annurev-astro-081817-051853}

\bibitem[{{Debes} \& {Sigurdsson}(2002)}]{DebesSigurdsson2002}
{Debes}, J.~H., \& {Sigurdsson}, S. 2002, \apj, 572, 556,
  \dodoi{10.1086/340291}

\bibitem[{{Duquennoy} \& {Mayor}(1991)}]{Duquennoy+91}
{Duquennoy}, A., \& {Mayor}, M. 1991, \aap, 248, 485

\bibitem[{{Eggleton} {et~al.}(1998){Eggleton}, {Kiseleva}, \& {Hut}}]{1998EKH}
{Eggleton}, P.~P., {Kiseleva}, L.~G., \& {Hut}, P. 1998, \apj, 499, 853,
  \dodoi{10.1086/305670}

\bibitem[{{El-Badry} \& {Rix}(2018)}]{ElBadryRix18}
{El-Badry}, K., \& {Rix}, H.-W. 2018, \mnras, 480, 4884,
  \dodoi{10.1093/mnras/sty2186}

\bibitem[{{Fabrycky} {et~al.}(2007){Fabrycky}, {Johnson}, \&
  {Goodman}}]{Fabrycky+07}
{Fabrycky}, D.~C., {Johnson}, E.~T., \& {Goodman}, J. 2007, \apj, 665, 754,
  \dodoi{10.1086/519075}

\bibitem[{{Fellhauer} {et~al.}(2003){Fellhauer}, {Lin}, {Bolte}, {Aarseth}, \&
  {Williams}}]{Fellhauer2003}
{Fellhauer}, M., {Lin}, D.~N.~C., {Bolte}, M., {Aarseth}, S.~J., \& {Williams},
  K.~A. 2003, \apjl, 595, L53, \dodoi{10.1086/379005}

\bibitem[{{Gettel} {et~al.}(2012){Gettel}, {Wolszczan}, {Niedzielski}, {Nowak},
  {Adam{\'o}w}, {Zieli{\'n}ski}, \& {Maciejewski}}]{Gettel+2012}
{Gettel}, S., {Wolszczan}, A., {Niedzielski}, A., {et~al.} 2012, \apj, 745, 28,
  \dodoi{10.1088/0004-637X/745/1/28}

\bibitem[{{Goldreich} \& {Tremaine}(1980)}]{GoldreichTremaine1980}
{Goldreich}, P., \& {Tremaine}, S. 1980, \apj, 241, 425, \dodoi{10.1086/158356}

\bibitem[{{Hamers}(2018)}]{Hamers18}
{Hamers}, A.~S. 2018, \mnras, 476, 4139, \dodoi{10.1093/mnras/sty428}

\bibitem[{{Hamilton} \& {Modak}(2024)}]{HamiltonModak2024}
{Hamilton}, C., \& {Modak}, S. 2024, \mnras, 532, 2425,
  \dodoi{10.1093/mnras/stae1654}

\bibitem[{{Hansen} \& {Naoz}(2020)}]{Hansen+20}
{Hansen}, B. M.~S., \& {Naoz}, S. 2020, \mnras, 499, 1682,
  \dodoi{10.1093/mnras/staa2602}

\bibitem[{Harris {et~al.}(2020)Harris, Millman, van~der Walt, Gommers,
  Virtanen, Cournapeau, Wieser, Taylor, Berg, Smith, Kern, Picus, Hoyer, van
  Kerkwijk, Brett, Haldane, del R{\'{i}}o, Wiebe, Peterson,
  G{\'{e}}rard-Marchant, Sheppard, Reddy, Weckesser, Abbasi, Gohlke, \&
  Oliphant}]{Harris+2020}
Harris, C.~R., Millman, K.~J., van~der Walt, S.~J., {et~al.} 2020, Nature, 585,
  357, \dodoi{10.1038/s41586-020-2649-2}

\bibitem[{{Heller}(2019)}]{Heller2019}
{Heller}, R. 2019, \aap, 628, A42, \dodoi{10.1051/0004-6361/201833486}

\bibitem[{{Heyl}(2008)}]{Heyl2008}
{Heyl}, J. 2008, \mnras, 390, 622, \dodoi{10.1111/j.1365-2966.2008.13724.x}

\bibitem[{{Hoang} {et~al.}(2022){Hoang}, {Naoz}, \& {Sloneker}}]{Hoang+22}
{Hoang}, B.-M., {Naoz}, S., \& {Sloneker}, M. 2022, \apj, 934, 54,
  \dodoi{10.3847/1538-4357/ac7787}

\bibitem[{{Holman} {et~al.}(1997){Holman}, {Touma}, \& {Tremaine}}]{Hol+97}
{Holman}, M., {Touma}, J., \& {Tremaine}, S. 1997, \nat, 386, 254,
  \dodoi{10.1038/386254a0}

\bibitem[{{Howard} {et~al.}(2012){Howard}, {Marcy}, {Bryson}, {Jenkins},
  {Rowe}, {Batalha}, {Borucki}, {Koch}, {Dunham}, {Gautier}, {Van Cleve},
  {Cochran}, {Latham}, {Lissauer}, {Torres}, {Brown}, {Gilliland}, {Buchhave},
  {Caldwell}, {Christensen-Dalsgaard}, {Ciardi}, {Fressin}, {Haas}, {Howell},
  {Kjeldsen}, {Seager}, {Rogers}, {Sasselov}, {Steffen}, {Basri},
  {Charbonneau}, {Christiansen}, {Clarke}, {Dupree}, {Fabrycky}, {Fischer},
  {Ford}, {Fortney}, {Tarter}, {Girouard}, {Holman}, {Johnson}, {Klaus},
  {Machalek}, {Moorhead}, {Morehead}, {Ragozzine}, {Tenenbaum}, {Twicken},
  {Quinn}, {Isaacson}, {Shporer}, {Lucas}, {Walkowicz}, {Welsh}, {Boss},
  {Devore}, {Gould}, {Smith}, {Morris}, {Prsa}, {Morton}, {Still}, {Thompson},
  {Mullally}, {Endl}, \& {MacQueen}}]{Howard+2012}
{Howard}, A.~W., {Marcy}, G.~W., {Bryson}, S.~T., {et~al.} 2012, \apjs, 201,
  15, \dodoi{10.1088/0067-0049/201/2/15}

\bibitem[{Hunter(2007)}]{Hunter2007}
Hunter, J.~D. 2007, Computing In Science \& Engineering, 9, 90,
  \dodoi{10.1109/MCSE.2007.55}

\bibitem[{{Hurley} {et~al.}(2000){Hurley}, {Pols}, \& {Tout}}]{Hurley+00}
{Hurley}, J.~R., {Pols}, O.~R., \& {Tout}, C.~A. 2000, \mnras, 315, 543,
  \dodoi{10.1046/j.1365-8711.2000.03426.x}

\bibitem[{{Hut}(1980)}]{Hut}
{Hut}, P. 1980, \aap, 92, 167

\bibitem[{{Innanen} {et~al.}(1997){Innanen}, {Zheng}, {Mikkola}, \&
  {Valtonen}}]{Innanen+1997}
{Innanen}, K.~A., {Zheng}, J.~Q., {Mikkola}, S., \& {Valtonen}, M.~J. 1997,
  \aj, 113, 1915, \dodoi{10.1086/118405}

\bibitem[{{Jackson} {et~al.}(2008){Jackson}, {Greenberg}, \&
  {Barnes}}]{Jackson+2008}
{Jackson}, B., {Greenberg}, R., \& {Barnes}, R. 2008, \apj, 678, 1396,
  \dodoi{10.1086/529187}

\bibitem[{{Johnson} {et~al.}(2011){Johnson}, {Payne}, {Howard}, {Clubb},
  {Ford}, {Bowler}, {Henry}, {Fischer}, {Marcy}, {Brewer}, {Schwab}, {Reffert},
  \& {Lowe}}]{Johnson+2011A}
{Johnson}, J.~A., {Payne}, M., {Howard}, A.~W., {et~al.} 2011, \aj, 141, 16,
  \dodoi{10.1088/0004-6256/141/1/16}

\bibitem[{{Jurado} {et~al.}(2024){Jurado}, {Naoz}, {Lam}, \&
  {Hoang}}]{Jurado+24}
{Jurado}, C., {Naoz}, S., {Lam}, C.~Y., \& {Hoang}, B.-M. 2024, \apj, 971, 95,
  \dodoi{10.3847/1538-4357/ad55ee}

\bibitem[{{Knutson} {et~al.}(2014){Knutson}, {Fulton}, {Montet}, {Kao}, {Ngo},
  {Howard}, {Crepp}, {Hinkley}, {Bakos}, {Batygin}, {Johnson}, {Morton}, \&
  {Muirhead}}]{Knutson+2014}
{Knutson}, H.~A., {Fulton}, B.~J., {Montet}, B.~T., {et~al.} 2014, \apj, 785,
  126, \dodoi{10.1088/0004-637X/785/2/126}

\bibitem[{{Kozai}(1962)}]{Kozai}
{Kozai}, Y. 1962, \aj, 67, 591, \dodoi{10.1086/108790}

\bibitem[{{Lagos} {et~al.}(2022){Lagos}, {Schreiber}, {Parsons}, {Toloza},
  {G{\"a}nsicke}, {Hernandez}, {Schmidtobreick}, \& {Belloni}}]{Lagos+2022}
{Lagos}, F., {Schreiber}, M.~R., {Parsons}, S.~G., {et~al.} 2022, \mnras, 512,
  2625, \dodoi{10.1093/mnras/stac673}

\bibitem[{{Li} {et~al.}(2014){Li}, {Naoz}, {Holman}, \& {Loeb}}]{Li+14}
{Li}, G., {Naoz}, S., {Holman}, M., \& {Loeb}, A. 2014, \apj, 791, 86,
  \dodoi{10.1088/0004-637X/791/2/86}

\bibitem[{{Lidov}(1962)}]{Lidov}
{Lidov}, M.~L. 1962, planss, 9, 719, \dodoi{10.1016/0032-0633(62)90129-0}

\bibitem[{{Lin} {et~al.}(1996){Lin}, {Bodenheimer}, \& {Richardson}}]{Lin+1996}
{Lin}, D.~N.~C., {Bodenheimer}, P., \& {Richardson}, D.~C. 1996, \nat, 380,
  606, \dodoi{10.1038/380606a0}

\bibitem[{{Lin} \& {Papaloizou}(1986)}]{Lin+86}
{Lin}, D.~N.~C., \& {Papaloizou}, J. 1986, \apj, 309, 846,
  \dodoi{10.1086/164653}

\bibitem[{{Lu} \& {Naoz}(2019)}]{LuNaoz2019}
{Lu}, C.~X., \& {Naoz}, S. 2019, \mnras, 484, 1506,
  \dodoi{10.1093/mnras/stz036}

\bibitem[{{Martin}(2018)}]{Martin2018}
{Martin}, D.~V. 2018, in Handbook of Exoplanets, ed. H.~J. {Deeg} \& J.~A.
  {Belmonte}, 156, \dodoi{10.1007/978-3-319-55333-7_156}

\bibitem[{{Martin} {et~al.}(2021){Martin}, {El-Badry}, {Hod{\v{z}}i{\'c}},
  {Triaud}, {Angus}, {Birky}, {Foreman-Mackey}, {Hedges}, {Montet}, {Murphy},
  {Santerne}, {Stassun}, {Stephan}, {Wang}, {Benni}, {Krushinsky}, {Chazov},
  {Mishevskiy}, {Ziegler}, {Soubkiou}, {Benkhaldoun}, {Boisse}, {Battley},
  {Miller}, {Caldwell}, {Collins}, {Henze}, {Guerrero}, {Jenkins}, {Latham},
  {Levine}, {McDermott}, {Mullally}, {Ricker}, {Seager}, {Shporer},
  {Vanderburg}, {Vanderspek}, \& {Winn}}]{Martin+21}
{Martin}, D.~V., {El-Badry}, K., {Hod{\v{z}}i{\'c}}, V.~K., {et~al.} 2021,
  \mnras, 507, 4132, \dodoi{10.1093/mnras/stab2129}

\bibitem[{{Masset} \& {Papaloizou}(2003)}]{Mass+03}
{Masset}, F.~S., \& {Papaloizou}, J.~C.~B. 2003, \apj, 588, 494,
  \dodoi{10.1086/373892}

\bibitem[{{Mazeh} \& {Shaham}(1979)}]{Mazeh+79}
{Mazeh}, T., \& {Shaham}, J. 1979, AA, 77, 145

\bibitem[{{Mendigut{\'\i}a} {et~al.}(2024){Mendigut{\'\i}a}, {Lillo-Box},
  {Vioque}, {Maldonado}, {Montesinos}, {Hu{\'e}lamo}, \&
  {Wang}}]{Mendigutia+24}
{Mendigut{\'\i}a}, I., {Lillo-Box}, J., {Vioque}, M., {et~al.} 2024, \aap, 686,
  L1, \dodoi{10.1051/0004-6361/202449368}

\bibitem[{{Michaely} \& {Naoz}(2023)}]{MichaelyNaoz2023}
{Michaely}, E., \& {Naoz}, S. 2023, arXiv e-prints, arXiv:2310.02558,
  \dodoi{10.48550/arXiv.2310.02558}

\bibitem[{{Michaely} \& {Shara}(2021)}]{MichaelyShara2021}
{Michaely}, E., \& {Shara}, M.~M. 2021, \mnras, 502, 4540,
  \dodoi{10.1093/mnras/stab339}

\bibitem[{{Moe} \& {Kratter}(2021)}]{MoeKratter21}
{Moe}, M., \& {Kratter}, K.~M. 2021, \mnras, 507, 3593,
  \dodoi{10.1093/mnras/stab2328}

\bibitem[{{Mu{\~n}oz} \& {Petrovich}(2020)}]{Munoz+2020}
{Mu{\~n}oz}, D.~J., \& {Petrovich}, C. 2020, arXiv e-prints, arXiv:2010.04724.
\newblock \doarXiv{2010.04724}

\bibitem[{{Nagasawa} {et~al.}(2008){Nagasawa}, {Ida}, \& {Bessho}}]{Nag+08}
{Nagasawa}, M., {Ida}, S., \& {Bessho}, T. 2008, \apj, 678, 498,
  \dodoi{10.1086/529369}

\bibitem[{{Naoz}(2016)}]{Naoz2016}
{Naoz}, S. 2016, \araa, 54, 441, \dodoi{10.1146/annurev-astro-081915-023315}

\bibitem[{{Naoz} {et~al.}(2011){Naoz}, {Farr}, {Lithwick}, {Rasio}, \&
  {Teyssandier}}]{Naoz11}
{Naoz}, S., {Farr}, W.~M., {Lithwick}, Y., {Rasio}, F.~A., \& {Teyssandier}, J.
  2011, \nat, 473, 187, \dodoi{10.1038/nature10076}

\bibitem[{{Naoz} {et~al.}(2013{\natexlab{a}}){Naoz}, {Farr}, {Lithwick},
  {Rasio}, \& {Teyssandier}}]{Naoz+11sec}
---. 2013{\natexlab{a}}, \mnras, 431, 2155, \dodoi{10.1093/mnras/stt302}

\bibitem[{{Naoz} {et~al.}(2012){Naoz}, {Farr}, \& {Rasio}}]{Naoz+12bin}
{Naoz}, S., {Farr}, W.~M., \& {Rasio}, F.~A. 2012, \apjl, 754, L36,
  \dodoi{10.1088/2041-8205/754/2/L36}

\bibitem[{{Naoz} {et~al.}(2016){Naoz}, {Fragos}, {Geller}, {Stephan}, \&
  {Rasio}}]{Naoz+2016}
{Naoz}, S., {Fragos}, T., {Geller}, A., {Stephan}, A.~P., \& {Rasio}, F.~A.
  2016, \apjl, 822, L24, \dodoi{10.3847/2041-8205/822/2/L24}

\bibitem[{{Naoz} {et~al.}(2013{\natexlab{b}}){Naoz}, {Kocsis}, {Loeb}, \&
  {Yunes}}]{Naoz+12GR}
{Naoz}, S., {Kocsis}, B., {Loeb}, A., \& {Yunes}, N. 2013{\natexlab{b}}, \apj,
  773, 187, \dodoi{10.1088/0004-637X/773/2/187}

\bibitem[{{Ngo} {et~al.}(2016){Ngo}, {Knutson}, {Hinkley}, {Bryan}, {Crepp},
  {Batygin}, {Crossfield}, {Hansen}, {Howard}, {Johnson}, {Mawet}, {Morton},
  {Muirhead}, \& {Wang}}]{Ngo+2016}
{Ngo}, H., {Knutson}, H.~A., {Hinkley}, S., {et~al.} 2016, \apj, 827, 8,
  \dodoi{10.3847/0004-637X/827/1/8}

\bibitem[{{O'Connor} {et~al.}(2020){O'Connor}, {Liu}, \& {Lai}}]{OConnor+2020}
{O'Connor}, C.~E., {Liu}, B., \& {Lai}, D. 2020, arXiv e-prints,
  arXiv:2010.04163.
\newblock \doarXiv{2010.04163}

\bibitem[{{Petrovich}(2015{\natexlab{a}})}]{Petrovich2015a}
{Petrovich}, C. 2015{\natexlab{a}}, \apj, 799, 27,
  \dodoi{10.1088/0004-637X/799/1/27}

\bibitem[{{Petrovich}(2015{\natexlab{b}})}]{Petrovich2015b}
---. 2015{\natexlab{b}}, \apj, 805, 75, \dodoi{10.1088/0004-637X/805/1/75}

\bibitem[{{Petrovich} \& {Mu{\~n}oz}(2017)}]{PetrovichMunoz2017}
{Petrovich}, C., \& {Mu{\~n}oz}, D.~J. 2017, \apj, 834, 116,
  \dodoi{10.3847/1538-4357/834/2/116}

\bibitem[{{Poon} {et~al.}(2021){Poon}, {Nelson}, \& {Coleman}}]{Poon+21}
{Poon}, S. T.~S., {Nelson}, R.~P., \& {Coleman}, G. A.~L. 2021, \mnras, 505,
  2500, \dodoi{10.1093/mnras/stab1466}

\bibitem[{{Raghavan} {et~al.}(2010){Raghavan}, {McAlister}, {Henry}, {Latham},
  {Marcy}, {Mason}, {Gies}, {White}, \& {ten Brummelaar}}]{Raghavan+10}
{Raghavan}, D., {McAlister}, H.~A., {Henry}, T.~J., {et~al.} 2010, \apjs, 190,
  1, \dodoi{10.1088/0067-0049/190/1/1}

\bibitem[{{Rasio} \& {Ford}(1996)}]{RasioFord1996}
{Rasio}, F.~A., \& {Ford}, E.~B. 1996, Science, 274, 954,
  \dodoi{10.1126/science.274.5289.954}

\bibitem[{{Raveh} {et~al.}(2022){Raveh}, {Michaely}, \& {Perets}}]{Yael+2022}
{Raveh}, Y., {Michaely}, E., \& {Perets}, H.~B. 2022, \mnras, 514, 4246,
  \dodoi{10.1093/mnras/stac1605}

\bibitem[{{Rein} \& {Liu}(2012)}]{ReinLiu2012}
{Rein}, H., \& {Liu}, S.~F. 2012, \aap, 537, A128,
  \dodoi{10.1051/0004-6361/201118085}

\bibitem[{{Rozner} \& {Perets}(2023)}]{MorPerets2023}
{Rozner}, M., \& {Perets}, H.~B. 2023, \apj, 955, 134,
  \dodoi{10.3847/1538-4357/ace2c6}

\bibitem[{{Shahaf} {et~al.}(2024){Shahaf}, {Hallakoun}, {Mazeh}, {Ben-Ami},
  {Rekhi}, {El-Badry}, \& {Toonen}}]{Shahaf+2024}
{Shahaf}, S., {Hallakoun}, N., {Mazeh}, T., {et~al.} 2024, \mnras, 529, 3729,
  \dodoi{10.1093/mnras/stae773}

\bibitem[{{Shariat} {et~al.}(2024){Shariat}, {Naoz}, {El-Badry}, {Rodriguez},
  {Hansen}, {Angelo}, \& {Stephan}}]{Shariat+2024}
{Shariat}, C., {Naoz}, S., {El-Badry}, K., {et~al.} 2024, arXiv e-prints,
  arXiv:2407.06257, \dodoi{10.48550/arXiv.2407.06257}

\bibitem[{{Shariat} {et~al.}(2023){Shariat}, {Naoz}, {Hansen}, {Angelo},
  {Michaely}, \& {Stephan}}]{Shariat+2023}
{Shariat}, C., {Naoz}, S., {Hansen}, B. M.~S., {et~al.} 2023, \apjl, 955, L14,
  \dodoi{10.3847/2041-8213/acf76b}

\bibitem[{{Spruit}(2002)}]{Spruit2002}
{Spruit}, H.~C. 2002, \aap, 381, 923, \dodoi{10.1051/0004-6361:20011465}

\bibitem[{{Stephan} {et~al.}(2018){Stephan}, {Naoz}, \& {Gaudi}}]{Stephan+2018}
{Stephan}, A.~P., {Naoz}, S., \& {Gaudi}, B.~S. 2018, \aj, 156, 128,
  \dodoi{10.3847/1538-3881/aad6e5}

\bibitem[{{Stephan} {et~al.}(2021){Stephan}, {Naoz}, \& {Gaudi}}]{Stephan+2021}
---. 2021, \apj, 922, 4, \dodoi{10.3847/1538-4357/ac22a9}

\bibitem[{{Stephan} {et~al.}(2016){Stephan}, {Naoz}, {Ghez}, {Witzel},
  {Sitarski}, {Do}, \& {Kocsis}}]{Stephan+2016}
{Stephan}, A.~P., {Naoz}, S., {Ghez}, A.~M., {et~al.} 2016, \mnras, 460, 3494,
  \dodoi{10.1093/mnras/stw1220}

\bibitem[{{Stephan} {et~al.}(2017){Stephan}, {Naoz}, \&
  {Zuckerman}}]{Stephan+2017}
{Stephan}, A.~P., {Naoz}, S., \& {Zuckerman}, B. 2017, \apjl, 844, L16,
  \dodoi{10.3847/2041-8213/aa7cf3}

\bibitem[{{Stephan} {et~al.}(2019){Stephan}, {Naoz}, {Ghez}, {Morris},
  {Ciurlo}, {Do}, {Breivik}, {Coughlin}, \& {Rodriguez}}]{Stephan+2019}
{Stephan}, A.~P., {Naoz}, S., {Ghez}, A.~M., {et~al.} 2019, \apj, 878, 58,
  \dodoi{10.3847/1538-4357/ab1e4d}

\bibitem[{{Tamayo} {et~al.}(2020){Tamayo}, {Rein}, {Shi}, \&
  {Hernandez}}]{Tamayo+2020}
{Tamayo}, D., {Rein}, H., {Shi}, P., \& {Hernandez}, D.~M. 2020, \mnras, 491,
  2885, \dodoi{10.1093/mnras/stz2870}

\bibitem[{{Teyssandier} {et~al.}(2019){Teyssandier}, {Lai}, \&
  {Vick}}]{Teyssandier+2019}
{Teyssandier}, J., {Lai}, D., \& {Vick}, M. 2019, \mnras, 486, 2265,
  \dodoi{10.1093/mnras/stz1011}

\bibitem[{{Triaud} {et~al.}(2010){Triaud}, {Collier Cameron}, {Queloz},
  {Anderson}, {Gillon}, {Hebb}, {Hellier}, {Loeillet}, {Maxted}, {Mayor},
  {Pepe}, {Pollacco}, {S{\'e}gransan}, {Smalley}, {Udry}, {West}, \&
  {Wheatley}}]{Tri+10}
{Triaud}, A.~H.~M.~J., {Collier Cameron}, A., {Queloz}, D., {et~al.} 2010,
  \aap, 524, A25+, \dodoi{10.1051/0004-6361/201014525}

\bibitem[{{Vanderburg} {et~al.}(2020){Vanderburg}, {Rappaport}, {Xu},
  {Crossfield}, {Becker}, {Gary}, {Murgas}, {Blouin}, {Kaye}, {Palle}, {Melis},
  {Morris}, {Kreidberg}, {Gorjian}, {Morley}, {Mann}, {Parviainen}, {Pearce},
  {Newton}, {Carrillo}, {Zuckerman}, {Nelson}, {Zeimann}, {Brown},
  {Tronsgaard}, {Klein}, {Ricker}, {Vand erspek}, {Latham}, {Seager}, {Winn},
  {Jenkins}, {Adams}, {Benneke}, {Berardo}, {Buchhave}, {Caldwell},
  {Christiansen}, {Collins}, {Col{\'o}n}, {Daylan}, {Doty}, {Doyle},
  {Dragomir}, {Dressing}, {Dufour}, {Fukui}, {Glidden}, {Guerrero}, {Guo},
  {Heng}, {Henriksen}, {Huang}, {Kaltenegger}, {Kane}, {Lewis}, {Lissauer},
  {Morales}, {Narita}, {Pepper}, {Rose}, {Smith}, {Stassun}, \&
  {Yu}}]{Vanderburg+2020}
{Vanderburg}, A., {Rappaport}, S.~A., {Xu}, S., {et~al.} 2020, \nat, 585, 363,
  \dodoi{10.1038/s41586-020-2713-y}

\bibitem[{{Veras}(2016)}]{Veras2016}
{Veras}, D. 2016, Royal Society Open Science, 3, 150571,
  \dodoi{10.1098/rsos.150571}

\bibitem[{{Veras} {et~al.}(2024){Veras}, {Mustill}, \& {Bonsor}}]{Veras+2024}
{Veras}, D., {Mustill}, A.~J., \& {Bonsor}, A. 2024, Reviews in Mineralogy and
  Geochemistry, 90, 141, \dodoi{10.2138/rmg.2024.90.05}

\bibitem[{{Veras} {et~al.}(2013){Veras}, {Mustill}, {Bonsor}, \&
  {Wyatt}}]{Veras+2013}
{Veras}, D., {Mustill}, A.~J., {Bonsor}, A., \& {Wyatt}, M.~C. 2013, \mnras,
  431, 1686, \dodoi{10.1093/mnras/stt289}

\bibitem[{{Virtanen} {et~al.}(2020){Virtanen}, {Gommers}, {Oliphant},
  {Haberland}, {Reddy}, {Cournapeau}, {Burovski}, {Peterson}, {Weckesser},
  {Bright}, {van der Walt}, {Brett}, {Wilson}, {Millman}, {Mayorov}, {Nelson},
  {Jones}, {Kern}, {Larson}, {Carey}, {Polat}, {Feng}, {Moore}, {VanderPlas},
  {Laxalde}, {Perktold}, {Cimrman}, {Henriksen}, {Quintero}, {Harris},
  {Archibald}, {Ribeiro}, {Pedregosa}, {van Mulbregt}, \& {SciPy 1. 0
  Contributors}}]{Virtanen+2020}
{Virtanen}, P., {Gommers}, R., {Oliphant}, T.~E., {et~al.} 2020, Nature
  Methods, 17, 261, \dodoi{10.1038/s41592-019-0686-2}

\bibitem[{{Weldon} {et~al.}(2024){Weldon}, {Naoz}, \& {Hansen}}]{Weldon+24}
{Weldon}, G.~C., {Naoz}, S., \& {Hansen}, B. M.~S. 2024, arXiv e-prints,
  arXiv:2405.20377, \dodoi{10.48550/arXiv.2405.20377}

\bibitem[{{Winn} {et~al.}(2010){Winn}, {Fabrycky}, {Albrecht}, \&
  {Johnson}}]{Winn+10b}
{Winn}, J.~N., {Fabrycky}, D., {Albrecht}, S., \& {Johnson}, J.~A. 2010, \apjl,
  718, L145, \dodoi{10.1088/2041-8205/718/2/L145}

\bibitem[{{Wolszczan} \& {Frail}(1992)}]{WolszczanFrail1992}
{Wolszczan}, A., \& {Frail}, D.~A. 1992, \nat, 355, 145,
  \dodoi{10.1038/355145a0}

\bibitem[{{Wu} {et~al.}(2023){Wu}, {Rice}, \& {Wang}}]{DongHongWu+2023}
{Wu}, D.-H., {Rice}, M., \& {Wang}, S. 2023, \aj, 165, 171,
  \dodoi{10.3847/1538-3881/acbf3f}

\bibitem[{{Wu} \& {Lithwick}(2011)}]{WuLithwick2011}
{Wu}, Y., \& {Lithwick}, Y. 2011, \apj, 735, 109,
  \dodoi{10.1088/0004-637X/735/2/109}

\bibitem[{{Wu} \& {Murray}(2003)}]{Wu+03}
{Wu}, Y., \& {Murray}, N. 2003, \apj, 589, 605, \dodoi{10.1086/374598}

\bibitem[{{Yamaguchi} {et~al.}(2024){Yamaguchi}, {El-Badry}, {Fuller},
  {Latham}, {Cargile}, {Mazeh}, {Shahaf}, {Bieryla}, {Buchhave}, \&
  {Hobson}}]{Yamaguchi+2024b}
{Yamaguchi}, N., {El-Badry}, K., {Fuller}, J., {et~al.} 2024, \mnras, 527,
  11719, \dodoi{10.1093/mnras/stad4005}

\bibitem[{{Zanazzi} \& {Wu}(2021)}]{ZanazziWu2021}
{Zanazzi}, J.~J., \& {Wu}, Y. 2021, \aj, 161, 263,
  \dodoi{10.3847/1538-3881/abf097}

\bibitem[{{Zink} \& {Howard}(2023)}]{ZinkHoward2023}
{Zink}, J.~K., \& {Howard}, A.~W. 2023, \apjl, 956, L29,
  \dodoi{10.3847/2041-8213/acfdab}

\end{thebibliography}

\end{CJK*}

\end{document}